\documentstyle[epsf]{article}
\begin{document}

\renewcommand{\baselinestretch}{1.5}
\large

\begin{center}
{\Large {\bf  An Attempt to Calculate Energy Eigenvalues \\
 in Quantum Systems of Large Sizes
} }
\vskip 0.8in
   T. MUNEHISA and Y.MUNEHISA
\vskip 0.2in
Faculty of Engineering, Yamanashi University

Takeda, Kofu, Yamanashi 400-8511, Japan
\vskip 0.5in
\end{center}
\vskip 0.5in
{\bf ABSTRACT}

We report an attempt to calculate energy eigenvalues of 
large quantum systems by the diagonalization of an effectively
truncated Hamiltonian matrix. 
For this purpose we employ a specific way to systematically make a set of
orthogonal states from a trial wavefunction and the Hamiltonian.
In comparison with the Lanczos method, which is quite powerful if the size 
of the system is within the memory capacity of computers, 
our method requires much less memory resources at the cost of the extreme 
accuracy. 

In this paper we demonstrate that our method works well  
in the systems of one-dimensional frustrated spins up to 48 sites, 
of bosons on a chain up to 32 sites 
and of fermions on a ladder up to 28 sites.
We will see this method enables us to study eigenvalues of these 
quantum systems within reasonable accuracy.

\eject

\noindent{\bf 1  \ \ Introduction }

So far a lot of methods have been developed to calculate eigenvalues 
of large size quantum systems. 
They can be brought under two chief categories,
the exact diagonalization methods\cite{dagetal} 
and the Monte Carlo approaches\cite{book1}.
In the former category, where the Lanczos method proves itself to be quite 
useful, one can expect strictly accurate results as long as the size of 
a system is small enough. 
The reason why the size of a system is limited in the Lanczos method
is that it consumes enormous memory resources to keep every component of 
the state.  
Within this category the density renormalization group method\cite{white} 
seems most 
prosperous to study systems of large sizes, but its effectivity strongly 
depends on both the system's properties and its dimensionality.
By means of the Monte Carlo methods, on the other hand, it is possible to 
study quantum systems with many sites.  
Instead it becomes necessary to deal with the statistical 
errors and in some cases the so-called negative sign problem would prevent us
to obtain meaningful results.
Another disadvantage of the methods is that the dynamical quantities are 
out of their reach. 

In this paper we propose a new method to obtain the eigenvalues in
quantum systems of large sizes, which is to be classified into the exact 
diagonalization approaches. 
A basic idea is that we calculate expectation values 
$$  \langle \Psi \mid \hat{O}_i \hat{H} \hat{O}_j \mid \Psi \rangle ,$$
with a choice of a trial wavefunction $\mid \Psi \rangle $ and 
a set of trial operators 
$ \{ \hat{O}_i \}$ aimed to demand much less memory than that required 
in the Lanczos method,  
so that the effective Hamiltonian matrix is diagonalized
within reasonable computer resources. 
A good property of our method is that it is applicable to 
various types of systems. 

In the next section we give a brief description of the method, 
giving general guidelines to 
find out the trial wavefunction and the trial operators in the systematic 
way. Concrete expressions for them will be given in the following sections
together with the results for the spin, the boson and the fermion systems.  
The final section is devoted to summary and discussion.

\vskip 3cm
\noindent{\bf 2  \ \ Method }

In this section we outline our method to calculate the
energy eigenvalues and their eigenstates, comparing it with the Lanczos method
if necessary.  

Let us consider to calculate energy eigenvalues using a finite number of 
elements of the system's Hamiltonian $\hat{H}$,
$$ H_{ij} = \langle \psi_i \mid \hat{H} \mid \psi_j \rangle, $$
where the states $\{ \psi_1, \psi_2, \cdots , \psi_n \}$ form a orthonormal set
$$  \langle \psi_i \mid \psi_j \rangle  = \delta_{ij}. $$
A good choice of the set is essential to obtain accurate eigenvalues 
using the presently available computer resources. 

In the Lanczos method, one begins with   
an orthogonal set made from an initial state $\mid \Psi \rangle$,
\begin{eqnarray} 
\{ \mid \Psi \rangle, \hat{H} \mid \Psi \rangle, \hat{H}^2 \mid \Psi \rangle,
\cdots , \hat{H}^{n-1} \mid \Psi \rangle \}, 
\label{eq:setL}
\end{eqnarray}
the value of $n$ being, empirically, of order $100$.
Then the matrix elements are given by the expectation values
$\langle \Psi \mid \hat{H}^{k} \mid \Psi \rangle$ where $k$ runs from 0 to 
$2n-1$. 
How the CPU time and the memory resources needed to calculate 
these expectation values increase with the system size $N$?
To find an answer to this question we
assume, for concreteness, that the number of components of each state 
vector grows as $p^N$ with some positive integer $p$ and that the Hamiltonian 
$\hat{H}$ is the sum of the partial Hamiltonians 
$\hat{h}_i $ $(i=1, \cdots , N)$,
\begin{eqnarray} 
 \hat{H}= \sum_{i=1}^{N} \hat{h}_i,
\label{eq:Hh}
\end{eqnarray}
where $\hat{h_i}$ denotes interactions between the site $i$ and the site $i+1$.
The CPU time to obtain the expectation values 
$\langle \Psi \mid H^k \mid \Psi \rangle$ would be of order $Nk$ 
provided that the time to calculate $\hat{h}_i \mid \phi \rangle$ is in the
same order for any intermediate state $\mid \phi \rangle$. It should be 
noted that the CPU time increases not {\em exponentially} but {\em linearly}
as $N$ increases.
The number of the components, on the contrary, will rapidly increase as 
$(pN)^{k/2}$ if each $\hat{h}_i$ produces a new state. Since one needs to 
keep {\em every} component of the states in the evaluation, the exhaustion 
of the memory resources would prevent one to enlarge the system size $N$. 

Now let us introduce our method. We would like to emphasize that this method 
needs much less memory resources 
compared with the Lanczos method and moderate, 
executable CPU time to obtain results on the eigenvalues and the eigenstates 
within acceptable numerical errors.
Instead of the set (\ref{eq:setL}) used in the Lanczos method we employ
\begin{eqnarray} 
\{ \hat O_0 \mid \Psi \rangle \equiv \mid \Psi \rangle , 
\hat O_1 \mid \Psi \rangle , 
\hat O_2 \mid \Psi \rangle \cdots , \hat O_{n-1} \mid \Psi \rangle \},
\label{eq:setO}
\end{eqnarray}
where these statevectors should be linearly independent with each 
others and every operator $\hat O_i$ should have the same 
symmetry as the Hamiltonian $\hat H$ has---the translational invariance, 
the conservation of the total spin, the conservation of the particle 
number and so on---so that any state generated from 
$\hat O_i$ keeps the related quantum numbers.
These requests are, however, too general to find out definite expressions of 
$\hat O_i$. 
Being strongly motivated by the success of the Lanczos method we additively 
demand that the expectation values 
$\langle \Psi \mid \hat H^k \mid \Psi \rangle $ can be, in principle, 
accurately evaluated with the set 
$ \{ \mid \Psi \rangle , \hat O_1 \mid \Psi \rangle , 
\hat O_2 \mid \Psi \rangle \cdots , \hat O_{n-1} \mid \Psi \rangle \}$. 
Thus we find a hint for a systematic choice of $\{ \hat O_i \}$  
in the expression obtained using (\ref{eq:Hh}),
\begin{eqnarray} 
\langle \Psi \mid \hat{H}^{k} \mid \Psi \rangle
           =\sum_{i_1=1}^N  \sum_{i_2=1}^N \cdots \sum_{i_k=1}^N
\langle \Psi \mid  \hat h_{i_1}  \hat h_{i_2} \cdots  \hat h_{i_k} 
\mid \Psi \rangle . 
\label{eq:f1}
\end{eqnarray}
Since the number of components one needs to calculate each  
$\langle \Psi \mid  \hat h_{i_1}  \hat h_{i_2} \cdots  
\hat h_{i_k}\mid \Psi \rangle$
is of the order of $p^{k/2}$, it seems favorable for our purpose to 
select the operators $\{ \hat O_i \}$
from $\{ \hat h_{i_1}  \hat h_{i_2} \cdots  \hat h_{i_k} \}$. 
Although we can deal with, of course, only a small portion of the whole 
$\{ \hat h_{i_1}  \hat h_{i_2} \cdots  \hat h_{i_k} \}$ for systems of 
large sizes because the number of terms in the sum increases as $N^k$,
we will see it is often the case that such $\{ \hat O_i \}$ is enough to 
obtain the satisfying results.  

As for the initial state $\mid \Psi \rangle$, the symmetry of the state
we want to study---the lowest energy state for example---is useful to 
determine it. It is desirable to choose a $\mid \Psi \rangle$ possessing 
the same quantum numbers as the state to be studied. 
The restructuring method\cite{munes} for re-arrangements of states is, 
in addition, quite helpful to get the $\mid \Psi \rangle$ which effectively 
reduces the CPU time in the calculation. 

Once the operators $\{ \hat O_i \}$ and the initial state 
$\mid \Psi \rangle$ are determined for the system under consideration,
the rest of the calculation is conventional. 
We calculate the Hamiltonian matrix $H_{Oij}$ $(i,j = 1, \cdots, n)$,
\begin{eqnarray}
H_{Oij} \equiv \langle \Psi \mid \hat O_{i-1} \hat{H} \hat O_{j-1} \mid \Psi 
\rangle . 
\label{eq:f2}
\end{eqnarray}
Then, employing the Gram-Schmidt method 
to create an orthonormal set \\ 
$\{ \mid \psi_1 \rangle,  \cdots , \mid ~ \psi_n \rangle \}$,
we obtain the coefficients $t_{ki}$,
$$ \mid \psi_k \rangle = \sum_{i=1}^k \hat{O}_{i-1} \mid \Psi \rangle t_{ki}, 
$$
which are utilized to calculate the effective Hamiltonian matrix 
$\overline{H}_{Oij}$,
\begin{eqnarray}  
\overline{H}_{Oij} =
 \sum_{i'=1}^n \sum_{j'=1}^n H_{Oi'j'}t_{ii'}t_{jj'} , 
\label{eq:f3}
\end{eqnarray}
where $t_{ki} \equiv 0$ for $i > k $.
The resultant $\overline{H}_{Oij}$ can be easily diagonalized by means of 
the traditional numerical methods.

\vskip 3cm
\noindent{\bf 3  \ \ Spin system }

In this section we apply our method to a quantum spin system.
We employ the one-dimensional frustrated system which 
has both the nearest-neighbor and the next-nearest-neighbor interactions, 
concentrating our attention to its ground state. 
The reasons why we adopt this system here are that its ground state is 
non-trivial and that the exact energy eigenvalues obtained by other methods
are available for the system of small sizes.

The Hamiltonian we study is
\begin{eqnarray}
\hat{H} = 
\sum_{i=1}^{N'} \vec{\sigma}_{a,i}\vec{\sigma}_{a,i+1}
+\sum_{i=1}^{N'} \vec{\sigma}_{b,i}\vec{\sigma}_{b,i+1}
+\sum_{i=1}^{N'} \vec{\sigma}_{a,i}\vec{\sigma}_{b,i}
+\sum_{i=1}^{N'} \vec{\sigma}_{b,i}\vec{\sigma}_{a,i+1},
\label{eq:Hs}
\end{eqnarray}
where the suffix $a$ ($b$) denotes the odd (even) sites on a
chain with $N \equiv 2N'$ spins and the periodic boundary condition 
$\vec{\sigma}_{a(b),N'+1} \equiv \vec{\sigma}_{a(b),1}$ is imposed.

First let us discuss on the initial state $\mid \Psi \rangle$.
In conventional calculations each state of the system is represented by
$\mid s_{a,1}, s_{b,1}, s_{a,2}, s_{b,2}, \cdots , s_{a,N'}, s_{b,N'} \rangle
$
with each $s_{a(b),i}= \pm$, 
which indicates that the $z$ component of the spin is $\pm 1/2$.
In the restructuring method\cite{munes} 
we rearrange states $s_{a,i}$ and $s_{b,i}$ into
a singlet state $ \mid \ominus_i \rangle $ and three triplet states
$ \mid 1_i \rangle$, $\mid \oplus_i \rangle$ and $\mid -1_i \rangle$,
\begin{eqnarray}
\nonumber
\mid \ominus _i\rangle & = & {1 \over \sqrt{2}} \left (
\mid +_{a,i},-_{b,i} \rangle - \mid -_{a,i},+_{b,i} \rangle \right ), \\
\nonumber
\mid 1_i \rangle & = & \mid +_{a,i}, +_{b,i} \rangle , \\
\nonumber
\mid \oplus _i \rangle & = & {1 \over \sqrt{2}} \left (
\mid +_{a,i},-_{b,i} \rangle + \mid -_{a,i},+_{b,i} \rangle \right ), \\
\mid -1_i \rangle & = & \mid -_{a,i}, -_{b,i} \rangle ,
\label{eq:Si}
\end{eqnarray}
which we inclusively denote by $\mid S_i \rangle$.  
The complete set for the states of the system is now given by 
$\{ \mid S_1, S_2, \cdots , S_{N'} \rangle \}$.
Since the spin-singlet state in this set is expressed by only one component
we employ this as the initial state, 
$$ \mid \Psi \rangle =\mid \ominus_1,\ominus_2, \cdots ,\ominus_{N'} \rangle, 
$$
so that we can largely reduce the CPU time needed in the calculation.
It should be minded
that the contamination of the states with momentum $\pi$  
is an undesirable feature of this choice, because the ground state should 
have the zero momentum only.
In spite of this fact, however, we will see we can obtain satisfying 
results with this $\mid \Psi \rangle$. 

Next we construct the operators $\{ \hat O_i \}$ from $\hat H^k$.
Taking the restructuring method into account, we resolve the Hamiltonian 
$\hat H$ into its components 
$\vec{\sigma}_{a,i}\vec{\sigma}_{b,i}$ and 
$\hat{h}_i \equiv \vec{\sigma}_{a,i}\vec{\sigma}_{a,i+1}
 +\vec{\sigma}_{b,i}\vec{\sigma}_{b,i+1}
 +\vec{\sigma}_{b,i}\vec{\sigma}_{a,i+1}$, 
$$  \hat{H} = \sum_{i=1}^{N'} \vec{\sigma}_{a,i}\vec{\sigma}_{b,i}
+ \sum_{i=1}^{N'} \hat{h}_{i} .
$$
Then in the expansion of $\hat H^k$ we direct our attention to the terms 
which are products of several $\hat{h}_{i}$'s 
with the suffix $i$ being in the whole range $1 \le i \le N'$.  
We combine those terms to express the candidates for $\{ \hat O_i \}$, 
which are described by
\begin{eqnarray}
\hat{O}(k_1, k_2, \cdots , k_{L}) \equiv 
\sum_{i=1}^{N'}\hat{h}_{i+k_1} \hat{h}_{i+k_2} \cdots  \hat{h}_{i+k_L}
\label{eq:Os}
\end{eqnarray}
with some positive integer $L$, which we will call 
{\it the order of the operator}
hereafter, and some positive integers $k_2$, $k_3$, $\cdots$, $k_L$.
Here we keep $k_1 \equiv 0$. 

Now we show the numerical results.
First let us report that this method presents the exact 
eigenvalues for the system of small sizes up to $N=12$.
For $N=8$ the set of the states
$$\{ \mid \Psi \rangle , \hat{O}_1 \mid \Psi \rangle , \cdots ,
 \hat{O}_5 \mid \Psi \rangle \}$$
with 
$$ \mid \Psi \rangle = \mid \ominus_1,\ominus_2, \ominus_3,\ominus_4 \rangle ,
$$
and
$$ \hat{O}_1=\hat{O}(0), \; \;
 \hat{O}_2=\hat{O}(0,1), \; \;
 \hat{O}_3=\hat{O}(0,2), \; \;
 \hat{O}_4=\hat{O}(0,1,0), \; \;
 \hat{O}_5=\hat{O}(0,1,2)$$
gives us the exact eigenvalues\footnote{Note that the number of 
the states necessary for the set also strongly depends on the system's 
property. 
For the quantum spin system on a ladder where there are no cross terms 
$ \vec{\sigma}_{b,i}\vec{\sigma}_{a,i+1}$, for instance, three states are 
enough to obtain the exact eigenvalue.}.
 
When $N=12$ the exact eigenvalue is obtainable with 28 states selected in 
the similar way. Table 1 presents the list of the operators $\hat O_i$ in 
this case. 
In order to see, without using other methods,    
whether the calculated eigenvalue of $\hat H$ is exact, 
we need to calculate the corresponding eigenvalue of $\hat H^2$, too.    
If the latter value agrees with the square of the former, 
the result is exact. 
Table 2 shows that the differences between the two for the $N=12$ chain are
negligible.

In the cases of larger sizes we have to do with approximate 
eigenvalues obtained from a limited number of the operators. To make a 
systematic selection of $\hat O_i $ from its candidates, we gradually 
increase the value of the order of the operator $L$ in Eq.(\ref{eq:Os}). 
Suppose that we have already picked up the operators up to, 
say, $L \le L_0$. The next step of the procedure is to list up  
$ \hat{O}(k_1,k_2, \cdots ,k_{L_0+1}) $'s with possible values of 
$\{ k_2, k_3, \cdots, k_{L_0+1} \}$ ($k_1 \equiv 0$) and add each of them 
into the set of $\{ \hat O_i \}$ if it creates a new state, namely a state 
which is linearly independent of all the previously chosen 
$\hat O_i \mid \Psi \rangle$'s, when operated to the initial state 
$\mid \Psi \rangle$. The operators in Table 1 also have been generated by this 
way. 
Note that the whole procedure is carried out by 
a computer program so that any work by hand is not necessary here.

In Table 3 we show the approximate eigenvalues of the ground state energy 
in the case of $N$=48 for $L \le 5$, together with 
the exact one obtained by the Lanczos method\cite{dagetal}. 
We see the approximation is improved as we pick up 
more operators bearing the cost of the longer CPU time, which is mainly 
originated from the increasing number of the states. 
In other words, in order to investigate systems of larger sizes within the 
available CPU time we have to adopt less precise results.
So the maximum size of the system is determined by arranging a compromise
between these factors. 
The size $N$=48 is the largest one we study in this paper.   
 
\vskip 3cm
\noindent{\bf 4  \ \ Boson system }

Now let us turn our attention to boson systems with the invariant boson 
number $N_b$. 
We consider the one-dimensional system with $N$ sites, whose Hamiltonian is 
given by 
\begin{eqnarray}
\hat{H} = -t \sum_{i=1}^{N} ( \hat{a}^{\dagger}_{i+1}\hat{a}_i 
 +  \hat{a}^{\dagger}_{i} \hat{a}_{i+1} )   + 
\lambda   \sum_{i=1}^{N} ( \hat{a}^{\dagger}_i  \hat{a}_i )^2,
\label{eq:Hb}
\end{eqnarray} 
$ \hat{a}_i(\hat{a}^{\dagger}_i)$ being an annihilation 
(a creation) operator on the site $i$, with the periodic boundary condition 
$\hat a^{(\dagger)}_{N+1} \equiv \hat a^{(\dagger)}_1$. In the numerical work
we take $t \equiv 1$, while $\lambda = 4$ unless stated 
otherwise.  

Limiting ourselves to the case $N_b=N$ we employ a translational 
invariant state $\mid 1, 1, \cdots, 1 \rangle$, where  
each site of the chain has one boson on it, as the initial state 
$\mid \Psi \rangle$. Note that we expect this is a good choice for 
large values of the coupling $\lambda$ since the state is the exact 
ground state in the strong coupling limit $\lambda \rightarrow \infty$. 

As for the operators $\{ \hat O_i \}$,
we follow the same procedure as previously stated in Section 3 to select them 
from the candidates
\begin{eqnarray}
\hat{O}(k_1,k_2, \cdots ,k_{L}) \equiv 
\sum_{i=1}^{N}\hat{g}_{i+k_1}
\hat{g}_{i+k_2} \cdots  \hat{g}_{i+k_L},
\label{eq:Ob}
\end{eqnarray}
where we use the notation $\hat g_{i} \equiv \hat{a}^{\dagger}_i  
\hat{a}_{i+1} +\hat{a}^{\dagger}_{i+1}  \hat{a}_i$.  
The reason why we do not include the diagonal terms of the Hamiltonian
in $\hat g_{i}$ is that more of the inner products 
$\langle \Psi \mid \hat O_i \hat O_j \mid \Psi \rangle$ will become zero 
without the diagonal terms.  

Figs.~1 and 2 present the results on the system's ground state energy. 
In Fig.~1 we plot the errors in our calculations for the $N=8$ system, 
which is small enough to obtain the exact value using the Lanczos method, 
as a function of the coupling $\lambda$.
The results 
indicate that our method works well in a wide range of the parameter space. 
Especially in the strong coupling region we see the agreement is excellent,
which confirms our expectation that the initial state we employed is 
good there.  
Fig.~2 shows our results on chains up to $N = 32$, 
together with the exact values for $N \le 10$.
The discrepancy between the ``exact'' value, which we guess from the 
exact values obtained on the short chains, 
and our result is about $0.5\%$ $(1\%)$ for
$N=16$ $(32)$, respectively.  These errors could be lessened  
by including the higher-order operators with $L \ge 5$.

\vskip 3cm
\noindent{\bf 5  \ \ Fermion system }

This section is devoted to the study of two fermion systems, the standard 
Hubbard model and the extended 
Hubbard model which has been extensively studied by the authors using 
a quantum Monte Carlo method\cite{ehm}.  Here we limit ourselves only 
to the ground state of the half-filled system. We will show our method  
nicely works for these systems, too.
  
The Hamiltonian of the extended Hubbard model we study is
\begin{eqnarray}
\nonumber
\hat{H} = -t_c
 \sum_{\sigma} \sum_{l} \sum_{i=1}^{N_r}
                       [c^\dagger_{i,l,\sigma}c_{i+1,l,\sigma} + h.c.]
             -t_r \sum_{\sigma} \sum_{i=1}^{N_r}
                       [c^\dagger_{i,a,\sigma}c_{i,b,\sigma} + h.c.] \\
             +V_c
      \sum_{\sigma, \sigma'} \sum_{l} \sum_{i=1}^{N_r-1}
                        n_{i,l,\sigma}n_{i+1,l,\sigma'}
             +V_r \sum_{\sigma, \sigma'} \sum_{i=1}^{N_r}
                        n_{i,a,\sigma}n_{i,b,\sigma'}       
             +U   \sum_{l} \sum_{i=1}^{N_r}
                        n_{i,l,\uparrow}n_{i,l,\downarrow},
\label{eq:Hf}
\end{eqnarray}
where $c^{(\dagger)}_{i,l,\sigma}$ denotes the annihilation (creation)
operator for an electron,
whose spin being $\sigma$ ($\uparrow$ or $\downarrow$),
located on the $i$-th rung along
the leg $l$ ($a$ or $b$) of a ladder which has $N_r$ rungs in total
and $n$ is the number operator 
$n_{i,l,\sigma} \equiv c^\dagger_{i,l,\sigma}c_{i,l,\sigma}$.
We assume the periodic boundary conditions  
$c^{(\dagger)}_{N_r+1,l,\sigma} \equiv c^{(\dagger)}_{1,l,\sigma}$.
The standard Hubbard model is described by setting parameters 
$V_r = V_c = 0$.

In the calculation we employ the restructuring method.
We use a complete set $\{ \mid \alpha \rangle \}$, where  
\begin{eqnarray}
\mid \alpha \rangle = \mid S_1, S_2, \cdots , S_{N_r} \rangle ,
\label{eq:cpls}
\end{eqnarray}
with sixteen states in Table~4 to denote the state 
$S_i$ on the $i$-th rung\footnote{
Parameters $u_1$ and $u_2$ in Table~4 are functions of $t_r$, $V_r$ and $U$
given by
$u_1 \equiv \frac{1}{2} \sqrt{1+ \frac{U-V_r}{\sqrt{(U-V_r)^2+16t_r^2}}}$ and
$u_2 \equiv \frac{1}{2} \sqrt{1- \frac{U-V_r}{\sqrt{(U-V_r)^2+16t_r^2}}}$.  }. 
It should be noted that the states in the table are eigenstates for
the interaction terms between two sites on the $i$-th rung, 
namely eigenstates for  
$$ -t_r \sum_\sigma [c^\dagger_{i,a,\sigma}c_{i,b,\sigma} + h.c.]
  +V_r \sum_{\sigma, \sigma'}  n_{i,a,\sigma}n_{i,b,\sigma'}
  +U \sum_l n_{i,l,\uparrow}n_{i,l,\downarrow} .
$$

The operator $\hat O_i$ is the product of 
$$\hat{f}_{i} \equiv -t_c \sum_{\sigma} \sum_{l}
 [c^{\dagger}_{i,l,\sigma}  c_{i+1,l,\sigma} + h.c.]
+V_c \sum_{\sigma, \sigma'} \sum_{l} 
 n_{i,l,\sigma}n_{i+1,l,\sigma'} ,
$$ 
which comes from the interaction terms along the chains,   
selected from the candidates
\begin{eqnarray}
\hat{O}(k_1,k_2, \cdots ,k_{L}) \equiv 
\sum_{i=1}^{N}\hat{f}_{i+k_1}
\hat{f}_{i+k_{2}} \cdots  \hat{f}_{i+k_{L}}.
\label{eq:Of}
\end{eqnarray}  

First let us report the results for the standard Hubbard model. 
In this case we choose as an initial state $\mid \Psi \rangle$
a translational invariant state whose all $S_i$ are No.~9 in Table~4. 
We plot the results on the ground state energy per site in Fig.~3 
as a function of the maximum value of $L$, $L_{max}$, for several values 
of $N (= 2N_r)$ up to 28.
Here the values of the active parameters in (\ref{eq:Hf}) are set to be  
$t_c = 1$, $t_r =2$ and $U = 4$, respectively.
We see that the values in Fig.~3 nicely converge 
into a value around $-1.269$.
We would like to emphasize that this value is compatible with our newly 
calculated quantum Monte Carlo result for the same Hamiltonian with the 
different boundary conditions. 
Although we have to accept the poor statistical accuracy for the 
almost half-filled\footnote{In the Monte Carlo study the number of 
the fermions can 
slightly fluctuate according to the given value of the chemical potential.}  
standard Hubbard model because of the negative-sign problem, 
we obtain the value $-1.26 \pm 0.02$  
at the inverse temperature $\beta = 10$ on   
an $N=32$ ladder with the open boundary conditions.

Next we show the results on the extended Hubbard model with parameters 
$t_c = 1$, $t_r =2$, $V_c = 2$, $V_r =-4$ and $U = 4$ in (\ref{eq:Hf}). 
In our previous 
quantum Monte Carlo study\cite{ehm} 
it turned out that the negative-sign problem 
is not serious for these values of the parameters in a wide range of the 
chemical potential and that the system indicates a signal to the 
phase separation.
Since we have learned from this study that the ground state near the 
half-filled system is abundant in both the hole pairs on the rungs and the 
doubly occupied sites, we choose the initial state in which the state 
on each rung $i$ is No.~1 (No.~16) in Table 4 for even (odd) $i$.
In Fig.~4 we plot the results for the ground state energy per site on 
the even $N_r$ ladders obtained with such initial states.
They also show nice convergence and are close to the exact value
$-2.202$ on the $N = 8 $ ladder with the periodic boundary conditions.
The quantum Monte Carlo result for the mostly half-filled case of the model 
is $-2.234 \pm 0.007$ (statistical error only) with $\beta = 5$ and $N = 16$ 
under the open boundary conditions.

\vskip 3cm
\noindent{\bf 6  \ \ Summary and Discussion}

In this paper we showed an attempt to calculate eigenvalues of the Hamiltonian
in various quantum systems of large sizes. 
In the calculation we start from a trial wavefunction, which we should 
choose carefully taking account of the symmetry of the state to be studied. 
Then we make a set of states by repeatedly applying some operators 
$\{ \hat{O}_i \}$
to the trial wavefunction. Our basic idea is to determine each $\hat{O}_i$
systematically using several terms in the Hamiltonian of the system under 
consideration. 
Then we select a limited number of states to construct an effective matrix 
of the Hamiltonian, whose eigenvalues are easily obtained  
by the conventional diagonalization methods.
In our calculation restrictions from the memory resources of computers are
relaxed, while the available CPU time sets bounds to the obtainable accuracy 
and the reachable maximum system size. 
As concrete examples we demonstrate the results for low-dimensional
spin, boson and fermion systems. We find the method works satisfactorily 
for all cases we studied.

In order to pursue further precise numerical results or to apply our method 
to larger systems including higher-dimensional ones, however, 
we need to achieve some more improvements.
One of them concerns with the size of the effective Hamiltonian matrix. 
Since we select states from $H^k \mid \Psi \rangle $, the number of 
the candidate states rapidly increases as either the power $k$ 
or the size of the system becomes large.
For the purposes stated above, therefore, we would be obliged to construct 
the larger effective Hamiltonian matrices for which serious restrictions 
again would come from the available memory resources.
So we should develop an effective method to treat large size matrices.
Another problem is that, because of the rapidly increasing CPU time, 
it is difficult to increase the order of the operators beyond the values 
we have used here.
A possible breakthrough for it would be to develop techniques to predict which 
of the states are linearly dependent.

\vskip 3cm
\noindent{\bf Acknowledgement}

We would like to thank Prof. Nishimori for his programs on the 
diagonalization.

\eject

\eject
\begin{table}[h]
\centering
\begin{tabular}{|c| c| c c c c c ||c|c| c c c c c|} \hline
No. & Order  & $k_1$ & $k_2$ & $k_3$ & $k_4$ & $k_5$ & No. & Order & $k_1$ & $k_2$ & $k_3$ & $k_4$ & $k_5$   \\ \hline \hline  
   1 &  0 &  --- &  --- &  --- &  --- &  --- & 15 &  4 &  0 &  1 &  0 &  5 &  ---\\  \hline
   2 &  1 &  0 &  --- &  --- &  --- &  --- & 16 &  4 &  0 &  1 &  2 &  1 &  ---\\  \hline
   3 &  2 &  0 &  1 &  --- &  --- &  --- & 17 &  4 &  0 &  1 &  2 &  3 &  ---\\  \hline
   4 &  2 &  0 &  2 &  --- &  --- &  --- & 18 &  4 &  0 &  1 &  2 &  4 &  ---\\  \hline
   5 &  2 &  0 &  3 &  --- &  --- &  --- & 19 &  4 &  0 &  1 &  3 &  2 &  ---\\  \hline
   6 &  3 &  0 &  1 &  0 &  --- &  --- & 20 &  4 &  0 &  1 &  3 &  4 &  ---\\  \hline
   7 &  3 &  0 &  1 &  2 &  --- &  --- & 21 &  4 &  0 &  2 &  1 &  4 &  ---\\  \hline
   8 &  3 &  0 &  1 &  3 &  --- &  --- & 22 &  5 &  0 &  1 &  0 &  2 &  3\\  \hline
   9 &  3 &  0 &  1 &  4 &  --- &  --- & 23 &  5 &  0 &  1 &  0 &  2 &  4\\  \hline
  10 &  3 &  0 &  2 &  1 &  --- &  --- & 24 &  5 &  0 &  1 &  0 &  3 &  2\\  \hline
  11 &  3 &  0 &  2 &  4 &  --- &  --- & 25 &  5 &  0 &  1 &  0 &  3 &  4\\  \hline
  12 &  4 &  0 &  1 &  0 &  2 &  --- & 26 &  5 &  0 &  1 &  2 &  1 &  4\\  \hline
  13 &  4 &  0 &  1 &  0 &  3 &  --- & 27 &  5 &  0 &  1 &  2 &  3 &  4\\  \hline
  14 &  4 &  0 &  1 &  0 &  4 &  --- & 28 &  5 &  0 &  2 &  1 &  4 &  3\\  \hline
\end{tabular}
\caption{ 
   Operators $\hat O (k_1, k_2, \cdots, k_L)$'s 
needed for the spin system on the $N=12$ chain. 
See (9) for the definition of the operator.
    }
\end{table}

\eject
\begin{table}[h]
\centering
\begin{tabular}{|c|c|c|c||c|c|c|c|} \hline
No. &  $\hat{H}$   & $\hat{H}^2$ & Diff. &
No. &  $\hat{H}$   & $\hat{H}^2$ & Diff.   \\ \hline \hline
   1 & $-$23.585     &  556.24     & 0.37064E-07& 15 & $-$7.0086     &  49.121     & 0.82701E-08\\  \hline
   2 & $-$22.901     &  524.44     & 0.37073E-07& 16 & $-$6.9468     &  48.258     & 0.25527E-07\\  \hline
   3 & $-$18.409     &  338.90     & 0.46285E-07& 17 & $-$4.2793     &  18.313     & 0.15361E-07\\  \hline
   4 & $-$17.271     &  298.29     & 0.27047E-07& 18 & $-$4.0000     &  16.000     & 0.10467E-07\\  \hline
   5 & $-$15.904     &  252.93     & 0.82821E-09& 19 & $-$3.4452     &  11.869     & 0.41185E-08\\  \hline
   6 & $-$13.904     &  193.31     & 0.29984E-07& 20 & $-$1.5279     &  2.3344     & 0.98762E-10\\  \hline
   7 & $-$13.411     &  179.85     & 0.44405E-07& 21 & $-$1.2330     &  1.5204     & 0.10958E-08\\  \hline
   8 & $-$12.000     &  144.00     & 0.26574E-07& 22 &$-$0.69308     & 0.48036     & 0.13380E-08\\  \hline
   9 & $-$11.446     &  131.00     & 0.42669E-07& 23 & 0.47600     & 0.22658     & 0.14751E-08\\  \hline
  10 & $-$11.266     &  126.92     & 0.33501E-08& 24 &  3.3595     &  11.286     & 0.52512E-08\\  \hline
  11 & $-$10.472     &  109.67     & 0.22215E-07& 25 &  4.8753     &  23.768     & 0.89002E-08\\  \hline
  12 & $-$10.065     &  101.29     & 0.31402E-09& 26 &  5.9206     &  35.054     & 0.53468E-09\\  \hline
  13 & $-$8.0485     &  64.778     & 0.26372E-07& 27 &  10.167     &  103.37     & 0.16889E-07\\  \hline
  14 & $-$7.6605     &  58.684     & 0.26150E-07& 28 &  16.676     &  278.10     & 0.32337E-07\\  \hline
\end{tabular}
\caption{
   Eigenvalues obtained in our calculations for the $N=12$ chain.
  Here the operators of the order $L\leq 5$ are used.
  The column of $\hat{H}$ shows the eigenvalues of the truncated
Hamiltonian matrix, while those from the Hamiltonian squared are shown in
the column of $\hat{H}^2$.
The third value is the difference between the second value and
the square of the first one.
    }
\end{table}

\eject
\begin{table}[h]
\centering
\begin{tabular}{|c| c|c| c | c|} \hline
Size &  Maximum order    & Num. of state & E  & E(exact) \\ \hline \hline
  16 & 4     &  37  & $-$1.935    &  $-$1.94010 \\  \hline
  20 & 4     & 59  &  $-$1.928 & $-$1.93565 \\  \hline
  24 & 4     &  92     & $-$1.920 & --- \\  \hline
  28 & 4     &  134     &  $-$1.911 & ---\\  \hline
  32 & 4     &  191     &  $-$1.902 & --- \\  \hline
  36 & 4     &  263     &  $-$1.892 & --- \\  \hline
  40 & 4     &  352    &  $-$1.883 & --- \\  \hline
  44 & 4     &  460    &  $-$1.874 & --- \\  \hline
  48 & 4     &  591     &  $-$1.865 & --- \\  \hline \hline
  12 & 5     &  28   & $-$1.96538  &  $-$1.96538 \\  \hline
 16 & 5     &  78  & $-$1.9040    &  $-$1.94010 \\  \hline
\end{tabular}
\caption{
   Eigenvalues of the ground state
  for the truncated Hamiltonian matrix up to $N \le 48$.
The second column (Maximum order) indicates the maximum value of $L$.
    }
\end{table}

\eject
\begin{table}[h]
\centering
\begin{tabular}{|r|c|}  \hline
  $ No. $ &  state \\ \hline
  $1$ & $\mid 00 \rangle $ \\ \hline
  $2$ & $\frac{1}{\sqrt{2}}(c^\dagger_{i,a,\uparrow} 
   +c^\dagger_{i,b,\uparrow})\mid 00 \rangle $ \\ \hline
  $3$ & $\frac{1}{\sqrt{2}}(c^\dagger_{i,a,\uparrow} 
   -c^\dagger_{i,b,\uparrow})\mid 00 \rangle $ \\ \hline
  $4$ & $\frac{1}{\sqrt{2}}(c^\dagger_{i,a,\downarrow} 
   +c^\dagger_{i,b,\downarrow})\mid 00 \rangle $ \\ \hline
  $5$ & $\frac{1}{\sqrt{2}}(c^\dagger_{i,a,\downarrow}
   -c^\dagger_{i,b,\downarrow})\mid 00 \rangle $ \\ \hline
  $6$ & $c^\dagger_{i,a,\uparrow}
         c^\dagger_{i,b,\uparrow}\mid 00 \rangle $ \\ \hline
  $7$ & $\frac{1}{\sqrt{2}}(c^\dagger_{i,a,\uparrow}
                            c^\dagger_{i,b,\downarrow}
                           +c^\dagger_{i,a,\downarrow}
         c^\dagger_{i,b,\uparrow}) \mid 00 \rangle $ \\ \hline
  $8$ & $\frac{1}{\sqrt{2}}(c^\dagger_{i,a,\uparrow}
                            c^\dagger_{i,a,\downarrow}
                          - c^\dagger_{i,b,\uparrow}
                   c^\dagger_{i,b,\downarrow}) \mid 00 \rangle $ \\ \hline
  $9$ & $[u_1 (c^\dagger_{i,a,\uparrow}
                            c^\dagger_{i,b,\downarrow}
         -c^\dagger_{i,a,\downarrow}
                           c^\dagger_{i,b,\uparrow})
        +u_2 (c^\dagger_{i,a,\uparrow}
                            c^\dagger_{i,a,\downarrow}
        + c^\dagger_{i,b,\uparrow}
                            c^\dagger_{i,b,\downarrow})]
\mid 00 \rangle $ \\ \hline
  $10$ & $[u_2 (c^\dagger_{i,a,\uparrow}
                            c^\dagger_{i,b,\downarrow}
         -c^\dagger_{i,a,\downarrow}
                           c^\dagger_{i,b,\uparrow})
        -u_1 (c^\dagger_{i,a,\uparrow}
                            c^\dagger_{i,a,\downarrow}
        + c^\dagger_{i,b,\uparrow}
                            c^\dagger_{i,b,\downarrow})]
\mid 00 \rangle $ \\ \hline
  $11$ & $c^\dagger_{i,a,\downarrow}
         c^\dagger_{i,b,\downarrow}\mid 00 \rangle $ \\ \hline
  $12$ & $\frac{1}{\sqrt{2}}
   (c^\dagger_{i,a,\uparrow} c^\dagger_{i,a,\downarrow}
    c^\dagger_{i,b,\uparrow}
   +c^\dagger_{i,a,\uparrow}
    c^\dagger_{i,b,\uparrow} c^\dagger_{i,b,\downarrow})
   \mid 00 \rangle $ \\ \hline
  $13$ & $\frac{1}{\sqrt{2}}
   (c^\dagger_{i,a,\uparrow} c^\dagger_{i,a,\downarrow}
    c^\dagger_{i,b,\uparrow}
   -c^\dagger_{i,a,\uparrow}
    c^\dagger_{i,b,\uparrow} c^\dagger_{i,b,\downarrow})
   \mid 00 \rangle $ \\ \hline
  $14$ & $\frac{1}{\sqrt{2}}
   (c^\dagger_{i,a,\uparrow} c^\dagger_{i,a,\downarrow}
    c^\dagger_{i,b,\downarrow}
   +c^\dagger_{i,a,\downarrow}
    c^\dagger_{i,b,\uparrow} c^\dagger_{i,b,\downarrow})
   \mid 00 \rangle $ \\ \hline
  $15$ & $\frac{1}{\sqrt{2}}
   (c^\dagger_{i,a,\uparrow} c^\dagger_{i,a,\downarrow}
    c^\dagger_{i,b,\downarrow}
   -c^\dagger_{i,a,\downarrow}
    c^\dagger_{i,b,\uparrow} c^\dagger_{i,b,\downarrow})
   \mid 00 \rangle $ \\ \hline
  $16$ & $ c^\dagger_{i,a,\uparrow} c^\dagger_{i,a,\downarrow}
           c^\dagger_{i,b,\uparrow} c^\dagger_{i,b,\downarrow}
   \mid 00 \rangle $ \\ \hline
\end{tabular}
\caption{States on two sites of the $i$-th rung 
         used to construct a complete set. $\mid 00 \rangle$ 
         represents a state with no electrons on either site 
         of the rung.
Parameters $u_1$ and $u_2$ are functions of $t_r$, $V_r$ and $U$ defined in 
Section 5.
  } 
\end{table}

\eject
\begin{figure}[hpt]
\centering
\epsfxsize=0.65\textwidth
\epsfbox{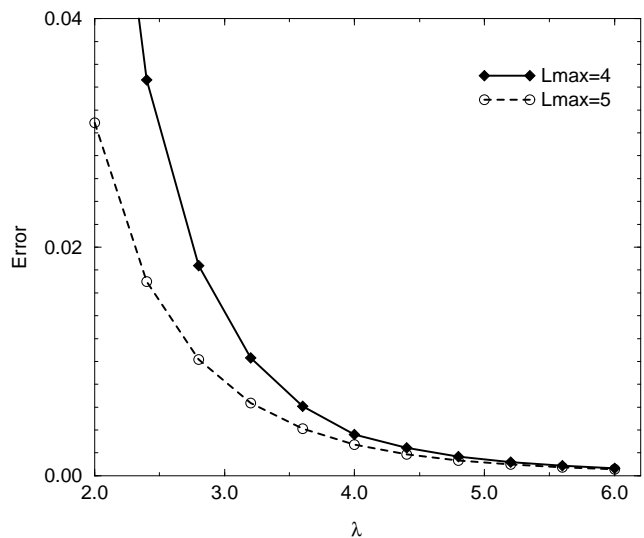}
\caption{Errors of the 
 calculated eigenvalues of the ground state plotted for several values of
the coupling $\lambda$ on the $N=8$ chain, which are defined by 
the ratio $( E_{cal} - E_{exact} ) / E_{exact}$.
 The diamonds show the errors for the order $L \le 4$, while
the circles are for the order $L \le 5$.}
\end{figure}

\begin{figure}[hpb]
\centering
\epsfxsize=0.65\textwidth
\epsfbox{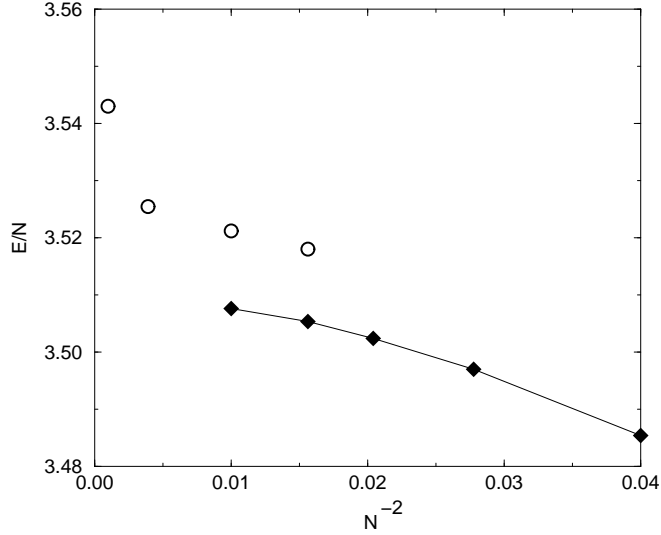}
\caption{
Eigenvalues of the ground state per site (the circles) 
calculated for $N = 8$, $10$, $16$ and $32$ chains  
with the coupling $\lambda = 4$ and the order $L \le 4$. 
The data are plotted versus $N^{-2}$.
The number of the states is 304 (1844) for $N=16$ (32), respectively.
Exact values for the systems up to $N \le 10$ are also shown by the 
diamonds.} 
\end{figure}

\eject
\begin{figure}[hpt]
\centering
\epsfxsize=0.65\textwidth
\epsfbox{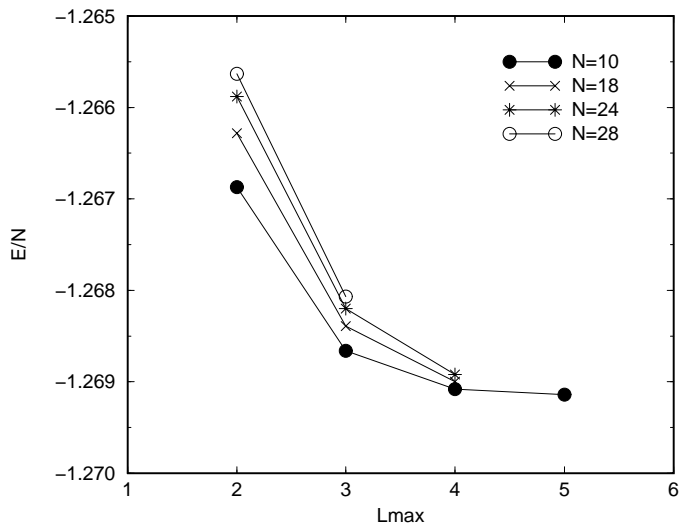}
\caption{Ground state energy per site for the standard
Hubbard model on ladders versus the maximum value of $L$, which is denoted by 
$L_{max}$, calculated for several values of the system size $N$. 
Lines in the figure are to guide eyes.
Values of the parameters in the Hamiltonian are $t_c=1$, $t_r=2$ and $U=4$.
 } 
\end{figure}

\begin{figure}[hpt]
\centering
\epsfxsize=0.65\textwidth
\epsfbox{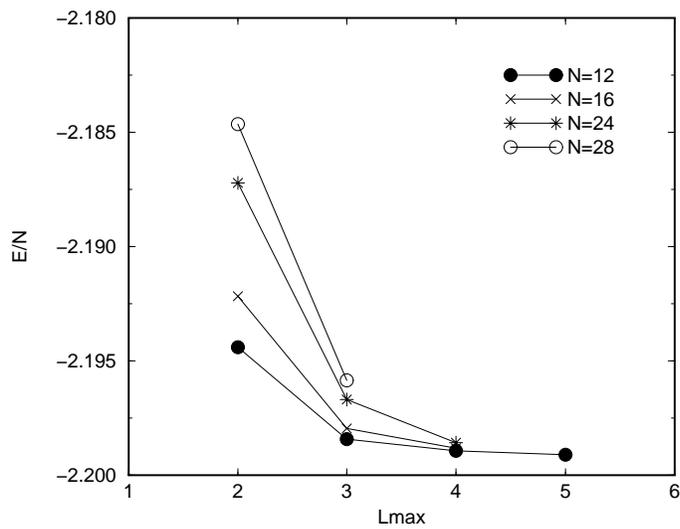}
\caption{Ground state energy per site for the extended Hubbard model 
described by (12) on ladders with size $N$ varying up to 28, 
as a function of $L_{max}$. 
Lines in the figure are to guide eyes.
Values of the parameters in the Hamiltonian are $t_c=1$, $t_r=2$, 
$V_c = 2$, $V_r =-4 $ and $U=4$.
 } 
\end{figure}

\end{document}